\documentclass[12pt]{iopart}

\usepackage{epsfig,amssymb}

\def\spose#1{\hbox to 0pt{#1\hss}}
\def\lesssim{\mathrel{\spose{\lower 3pt\hbox{$\mathchar"218$}}
 \raise 2.0pt\hbox{$\mathchar"13C$}}}
\def\gtrsim{\mathrel{\spose{\lower 3pt\hbox{$\mathchar"218$}}
 \raise 2.0pt\hbox{$\mathchar"13E$}}}

\begin{document}
\title{The random-anisotropy model in the strong-anisotropy limit}
\author{
Francesco Parisen Toldin,$^{1}$ 
Andrea Pelissetto,$^2$ and Ettore Vicari$^3$ 
}
\address{$^1$ Scuola Normale Superiore and INFN, Pisa, Italy}
\address{$^2$ Dipartimento di Fisica dell'Universit\`a di Roma 
``La Sapienza" and INFN, Roma, Italy.}
\address{$^3$ Dipartimento di Fisica dell'Universit\`a di Pisa 
and INFN, Pisa, Italy.
}
\ead{
{\tt f.parisentoldin@sns.it},
{\tt Andrea.Pelissetto@roma1.infn.it},
{\tt Ettore.Vicari@df.unipi.it}
}

\date{\today}

\begin{abstract}
    We investigate the nature of the critical behaviour of the random-anisotropy
    Heisenberg model (RAM), which describes a magnetic system with random
    uniaxial single-site anisotropy, such as some amorphous alloys of
    rare earths and transition metals. In particular, we consider the
    strong-anisotropy limit (SRAM), in which the Hamiltonian can be
    rewritten as the one of an Ising spin-glass model with correlated bond
    disorder: ${\cal H} = - J \sum_{\langle xy \rangle} j_{xy} \sigma_x
    \sigma_y$, where $j_{xy}= \vec{u}_x \cdot \vec{u}_y$ and $\vec{u}_x$ is
    a random three-component unit vector. We performed Monte Carlo
    simulations of the SRAM on simple cubic $L^3$ lattices, up to $L=30$,
    measuring correlation functions of the replica-replica overlap,
    which is the order parameter at a glass transition. The corresponding
    results show critical behaviour and finite-size scaling. They provide
    evidence of a finite-temperature continuous transition with critical
    exponents $\eta_o=-0.24(4)$ and $\nu_o=2.4(6)$. These results are
    close to the corresponding estimates that have been obtained in
    the usual Ising spin-glass model with uncorrelated bond disorder,
    suggesting that the two models belong to the same universality class.
    This is consistent with arguments that suggest that the disorder
    correlations present in the SRAM are irrelevant.
\end{abstract}


\section{Introduction}
\label{secint}

Extensive theoretical and experimental work has been devoted to the study of
amorphous alloys of rare earths and transition metals, for instance TbFe$_2$
and YFe$_2$. They are modeled \cite{HPZ-73} by a Heisenberg Hamiltonian
with random uniaxial single-site anisotropy defined on a simple cubic lattice,
or, in short, by the random-anisotropy model (RAM)
\begin{equation}
{\cal H} =- J \sum_{\langle xy \rangle} \vec{s}_x \cdot \vec{s}_y-
D \sum_x(\vec{u}_x \cdot \vec{s}_x)^2,
\label{HRAM}
\end{equation}
where $\vec{s}_x$ is a three-component spin variable, $\vec{u}_x$ is a unit
vector describing the local (spatially uncorrelated) random anisotropy, and
$D$ the anisotropy strength. In amorphous alloys the distribution of
$\vec{u}_x$ is isotropic, since, in the absence of
crystalline order, there is no preferred direction.  

Random anisotropy is a relevant perturbation of the pure Heisenberg model, the
crossover exponent being \cite{CPV-04} $\phi_D=0.412(3)$.  Therefore,
random-anisotropy systems show a behavior that is different from that observed
in pure Heisenberg systems. Even though the critical behavior of the RAM has
been investigated at length in the last thirty years (see \cite{DFH-05} for a
review) we do not have yet a satisfactory picture of its critical behaviour.
The Imry-Ma argument \cite{IM-73} forbids the presence of a low-temperature
phase with nonvanishing magnetization for $d<4$.  However, this does not
exclude the appearance of a glass transition with a low-temperature phase
characterized by quasi-long-range order (QLRO), i.e., a phase in which
correlation functions decay algebraically \cite{PPR-78}.  Functional
renormalization-group calculations \cite{Feldman-00,Feldman-01} indicate that
QLRO may set in for small values of $D$, in agreement with a
Landau-Ginzburg calculation \cite{AP-80} of the equation of state for $D\to
0$.  In the opposite limit $D\to \infty$ the RAM Hamiltonian reduces to that
of an Ising spin glass with a correlated bond distribution.  Indeed, for
$D\to\infty$ one can write $\vec{s}_x=\sigma_x \vec{u}_x$, where $\sigma_x=\pm
1$ is an Ising spin, and obtain the Hamiltonian
\begin{equation}
{\cal H} = - \sum_{\langle xy \rangle} j_{xy} \sigma_x \sigma_y,
\qquad j_{xy}\equiv  \vec{u}_x \cdot \vec{u}_y.
\label{raminfD} 
\end{equation}
We call this model strong random-anisotropy model (SRAM) (We set $J=1$ without
loss of generality).
Model (\ref{raminfD}) differs from the usual Ising spin glass model (ISGM) 
in the bond distribution: in Hamiltonian (\ref{raminfD})
the random variables $j_{xy}$ on different lattice links are 
correlated. For instance, one has 
$\overline{\prod_\square j_{xy}} = 1/27$, 
where the product is over the links belonging to a given 
plaquette and the average is taken with respect
to the distribution of the vectors $\vec{u}_x$.

An interesting hypothesis, originally put forward in \cite{CL-77}, is that the
SRAM transition is in the same universality class as that of the ISGM. This
conjecture looks very plausible, since the SRAM is nothing but an Ising model
with local disorder and frustration. In some sense we can think of the SRAM
and of the ISGM as two different versions of the same model: in the SRAM
disorder is associated with lattice sites, while in the ISGM disorder is
associated with lattice bonds.  They are analogous to the site-diluted and
bond-diluted Ising model \cite{PV-rev,FHY-03}, whose Hamiltonian is given by
(\ref{raminfD}) with $j_{xy} = r_x r_y$ (site dilution) and $j_{xy} = r_{xy}$
(bond dilution), $r$ being a random variable such that $r = 1$ ($r=0$) with
probability $p$ (resp. $1-p$). Note that the site-diluted model can also be
interpreted as a bond-diluted model with a correlated bond distribution,
exactly as is the case for the SRAM. Nonetheless, there is little
doubt---though a satisfactory numerical check is still missing---that the two
models belong to the same universality class.

Note that the SRAM is less frustrated than the standard ISGM, since in the
SRAM $\overline{\prod_\square j_{xy}} = 1/27$.  This fact does not rule out
the conjecture since it is known that maximal frustration is not necessary to
obtain glassy behaviour. For instance, the random-bond Ising model with
$j_{xy} = +1$ with probability $p$ and $j_{xy} = -1$ with probability $1-p$
has a glassy low-temperature phase for \cite{OI-98} $0.233 \lesssim p \lesssim
0.767$.

The identity of the SRAM and ISGM universality classes was confirmed in two
dimensions by a renormalization-group calculation using the large-cell method
\cite{BM-85}, though in this case the critical point is at $T=0$.  In three
dimensions instead, the critical behaviour of the SRAM has been controversial
for a long time. While, for small $D$, numerical simulations
\cite{JK-80,Chakrabarti-87,Fisch-90,Fisch-98,Itakura-03} provided some
evidence of the existence of a finite-temperature transition (though QLRO was
never observed), in the SRAM even the existence of a finite-temperature
transition was in doubt \cite{Itakura-03}.

In \cite{randa} we study again the SRAM and find good evidence for the
existence of a finite-temperature transition. The corresponding critical
behaviour turns out to be compatible with the conjecture of \cite{CL-77}.
  Note that at the transition
only the overlap variables, which are the usual order parameters at a
spin-glass transition, become critical.  Magnetic quantities do not show any
critical behaviour, though we expect nonanalyticities induced by the critical
modes.  Thus, on both sides of the transition the system is paramagnetic. It
is unknown whether this paramagnetic phase survives up to $T=0$.

Note that our results predict an ISGM transition also in 
a generalized SRAM in which the vectors $\vec{u}$ are $N$ dimensional, 
for $N> 3$.
Indeed, as $N$ increases, the bond correlation decreases and, for $N\to \infty$,
one reobtains the ISGM although with a nonstandard bond distribution. 
For $N=2$ instead a {\em ferromagnetic} phase transition would be possible
since the system is less frustrated than the case we have considered.
Note that such a transition would only be observed in 
correlations of $\epsilon_x \sigma_x$, where the Ising variables 
$\epsilon_x$ should be chosen such as to have the couplings 
$\epsilon_x \epsilon_y j_{xy}$ ferromagnetic on a maximal set of 
lattice links. This ferromagnetic transition 
would not violate the Imry-Ma argument, since 
order in the $\sigma_x$ variables does not imply order 
in the continuous variables $\vec{s}_x$. Most likely, as for $N=3$, the 
low-temperature phase would still be paramagnetic even if some Ising variables
magnetize. 

A question that remains open is the behaviour of the RAM for finite
anisotropy $D$. If there is indeed a low-temperature phase with
QLRO for small $D$ as predicted in \cite{Feldman-00,Feldman-01,AP-80},
then there should be a critical value $D^*$ such that ISGM behaviour
is observed only for $D>D^*$. Nothing is known about $D^*$ 
and we cannot even exclude that $D^*=\infty$, so that ISGM behaviour 
is observed only for model (\ref{raminfD}).

\section{Results}

In Ref.~\cite{randa}
we study the critical behaviour of the SRAM 
by means of Monte Carlo (MC) simulations. Since the model is 
essentially a spin glass we focus on the so-called overlap 
variables $\sigma_x\tau_x$, where $\sigma_x$ and $\tau_x$
are associated with two different replicas of the model with the same 
bond variables. For the SRAM one can also consider the 
standard magnetic variables $\vec{s}_x = \sigma_x \vec{u}_x$.
We find, in agreement with the results of \cite{Itakura-03},
that these quantities are not critical, i.e. on the low-temperature 
side of the transition the system is still paramagnetic.
We study the behaviour of the SRAM in the high-temperature 
phase. This reduces the algorithmic problems---the MC algorithm
becomes very slow as temperature is reduced---and allows us to consider 
lattices of size $L^3$ up to $L = 30$. For the MC dynamics,
we combine the Metropolis algorithm with the random-exchange
(parallel-tempering) method \cite{Geyer-91,HN-96}. 

\begin{figure}[tbp]
  \begin{minipage}[t]{0.5\linewidth}
    \includegraphics[width=0.9\textwidth, keepaspectratio]{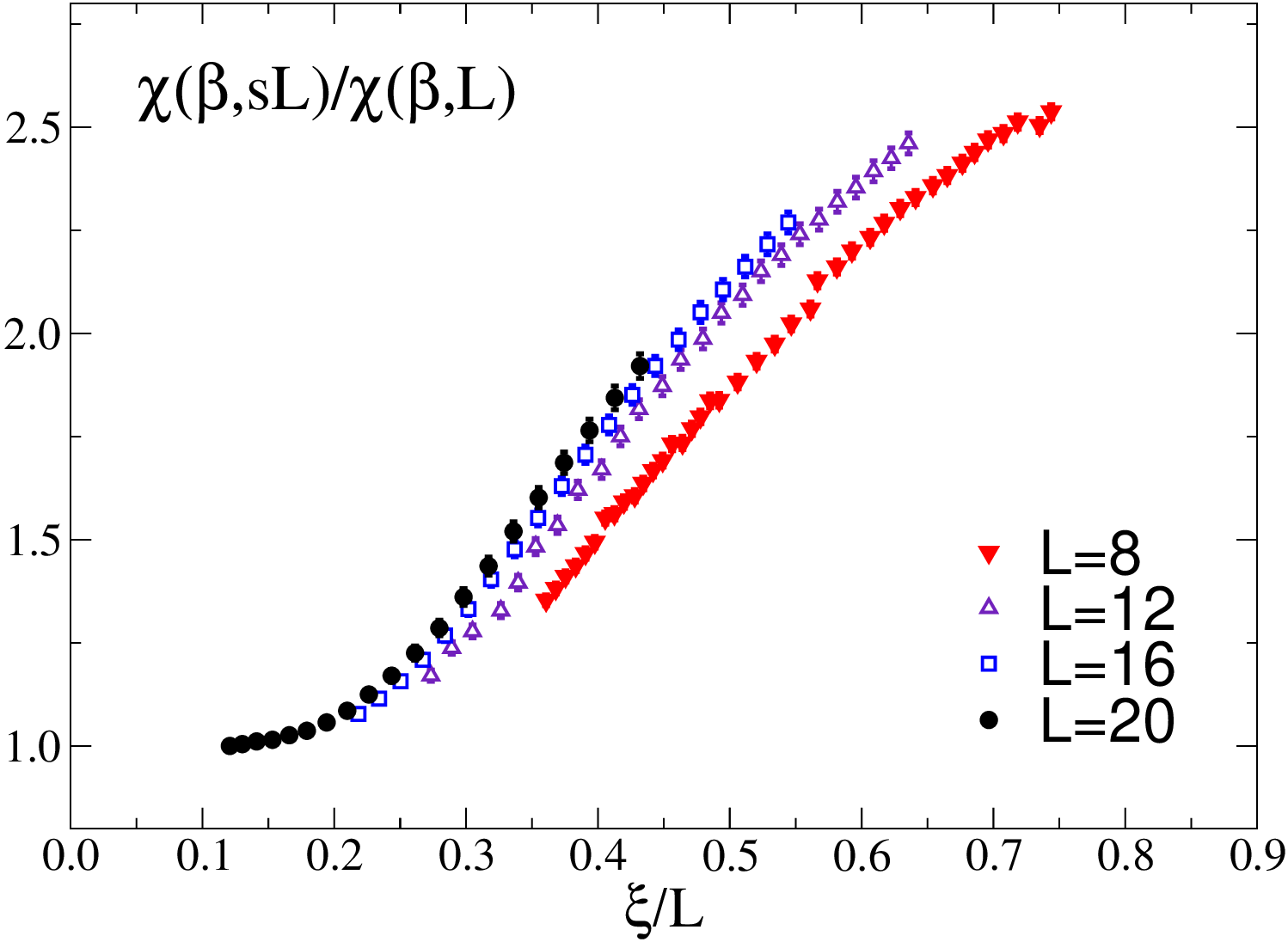}
    \setlength{\mathindent}{1em}
    \caption{The FSS curve of the susceptibility $\chi$ for $s=3/2$.}
    \label{chi}
  \end{minipage}
  \begin{minipage}[t]{0.5\linewidth}
    \includegraphics[width=0.9\textwidth, keepaspectratio]{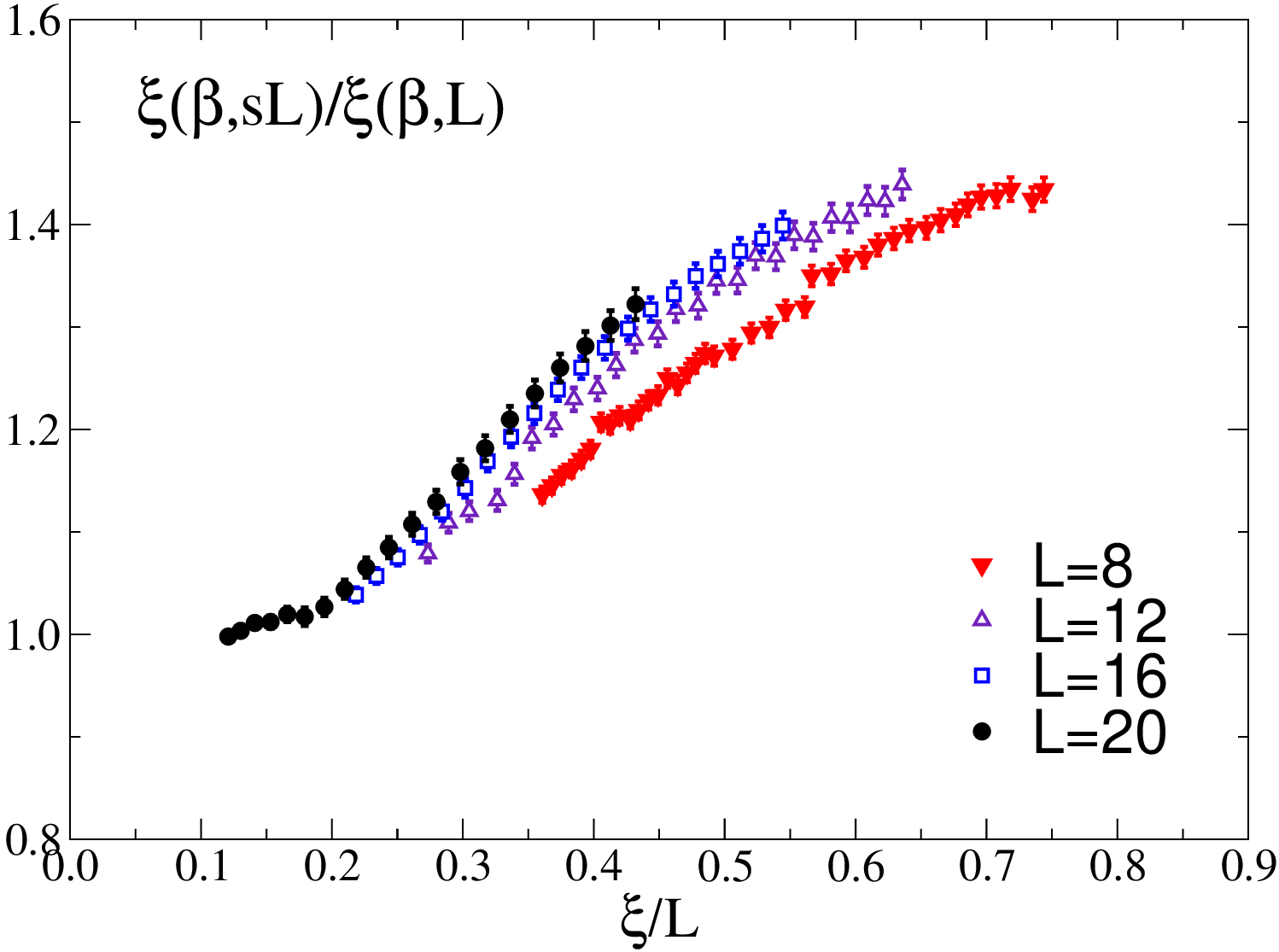}
    \setlength{\mathindent}{1em}
    \caption{The FSS curve of the correlation length $\xi$ for $s=3/2$.}
    \label{xi}
  \end{minipage}
\end{figure}

In order to verify whether the system has a critical behaviour,
we looked for the occurence of finite-size scaling (FSS).
For this purpose, we considered
the ratios $R_A\equiv A(\beta,sL)/A(\beta,L)$ with $A=\chi$, $\xi$
(both quantities are associated with the two-point function of the 
overlap variables),
fixing $s = 3/2$. If FSS holds,  as
$L$ increases all points should eventually
fall on a universal curve depending on $\xi(\beta,L)/L$.
Our results, reported in Figures \ref{chi}, \ref{xi},
show that this happens: 
The data corresponding to the pairs $L=16$, $24$ are
only slightly different from those with $L=20$, $30$, the 
difference being much less than that observed by comparing 
data with $L=16,24$ and $L=8,12$.
Therefore, the data strongly suggest that FSS holds, (though,
for $L\lesssim 30$, scaling corrections are significant) and therefore,
that the system becomes eventually critical.
At the critical point $R_\xi(\beta_c,L) = s$. Looking at Fig.~\ref{xi}
we see that all MC data satisfy $R_\xi(\beta,L)\lesssim 3/2 = s$, which 
indicates that they all lie in the high-temperature phase. 
This allows us to set a lower bound on the position of the critical 
point, $\beta_c\gtrsim 1.1$.

In order to determine the critical properties of the model,
i.e. critical point and critical exponents, we used
the iterative method which was introduced in \cite{CEFPS-95} and 
generalized in \cite{CMPV-03} to include scaling corrections.
It allowed us to obtain infinite-volume estimates of 
$\chi$ and $\xi$ up to
$\xi_\infty \approx 20$ ($\xi_\infty$ is the infinite-volume 
second-moment correlation length) in the high-temperature phase. 
The starting point is the FSS relation
\begin{equation}
{A(\beta,sL)\over A(\beta,L)} = 
  F_A(s,\xi(\beta,L)/L) + L^{-\omega} G_A(s,\xi(\beta,L)/L),
\label{FSS-corr}
\end{equation}
valid for any long-distance quantity. Here $s$ is an arbitrary number 
(in our calculations we fixed $s=3/2$), 
$F_A(s,z)$ and $G_A(s,z)$ are universal scaling functions, and 
$\omega$ is a to-be-determined correction-to-scaling exponent.
By iterating (\ref{FSS-corr}) it is possible to extrapolate 
the finite-volume estimates
$A(\beta, L)$, obtaining the infinite-volume
value $A_\infty(\beta)$.

Our results show that
$\xi_\infty(\beta)$ increases quite rapidly in the 
range we have studied, $0.80\lesssim \beta \lesssim 1.0$, 
confirming that the system eventually becomes critical.
Fitting the infinite-volume results,
we are able to determine the critical point
and critical exponents associated
with the critical behaviour of the overlap observables.
We obtain:
\begin{eqnarray}
   && \beta_c = 1.08(4) \nonumber \\
   && \eta_o = - 0.24(4) \nonumber\\
\label{results}
   && \nu_o = 2.4(6) \\
   && \gamma_o = 5.3(1.3). \nonumber
\end{eqnarray}
The critical exponents are defined by $\chi \sim \xi^{2- \eta_o}$ and 
$\xi \sim (\beta_c - \beta)^{-\nu_o}$.
The suffix $o$ is introduced to 
remind that they refer to the overlap variables and not to the 
magnetic ones. 
In the analysis it is crucial to include corrections
to scaling in the FSS (corrections behave as 
$L^{-\omega}$) and in fits of infinite-volume 
quantities (corrections behave as $\xi_\infty^{-\omega}$ as 
$\xi_\infty\to\infty$). 
The exponent
$\omega$ was determined by studying the critical behaviour of two 
universal ratios that involve the four-point and the two-point correlation
function of the overlap variables. We obtained
$\omega = 1.0 \pm 0.4$.

Estimates (\ref{results}) are close to those obtained for the ISGM and thus
support the conjecture that the SRAM transition is in the same universality
class as that of the ISGM.  Ref.~\cite{Joerg-06} quotes $\nu_o = 2.22(15)$,
$\eta_o = -0.349(18)$, while \cite{KKY-06} reports $\nu_o = 2.39(5)$ and
$\eta_o = -0.395(17)$.  Other estimates are reported in Table 1 of
\cite{KKY-06}.  With the quoted error bars there is a small discrepancy
between our estimate of $\eta_o$ and those of \cite{Joerg-06,KKY-06}. This
difference should not be taken too seriously, since there are similar
discrepancies among the estimates obtained by different groups for the bimodal
ISGM, see Table 1 of \cite{KKY-06}.  

\begin{figure}[tbp]
  \begin{minipage}[t]{0.5\linewidth}
    \includegraphics[width=0.9\textwidth, keepaspectratio]{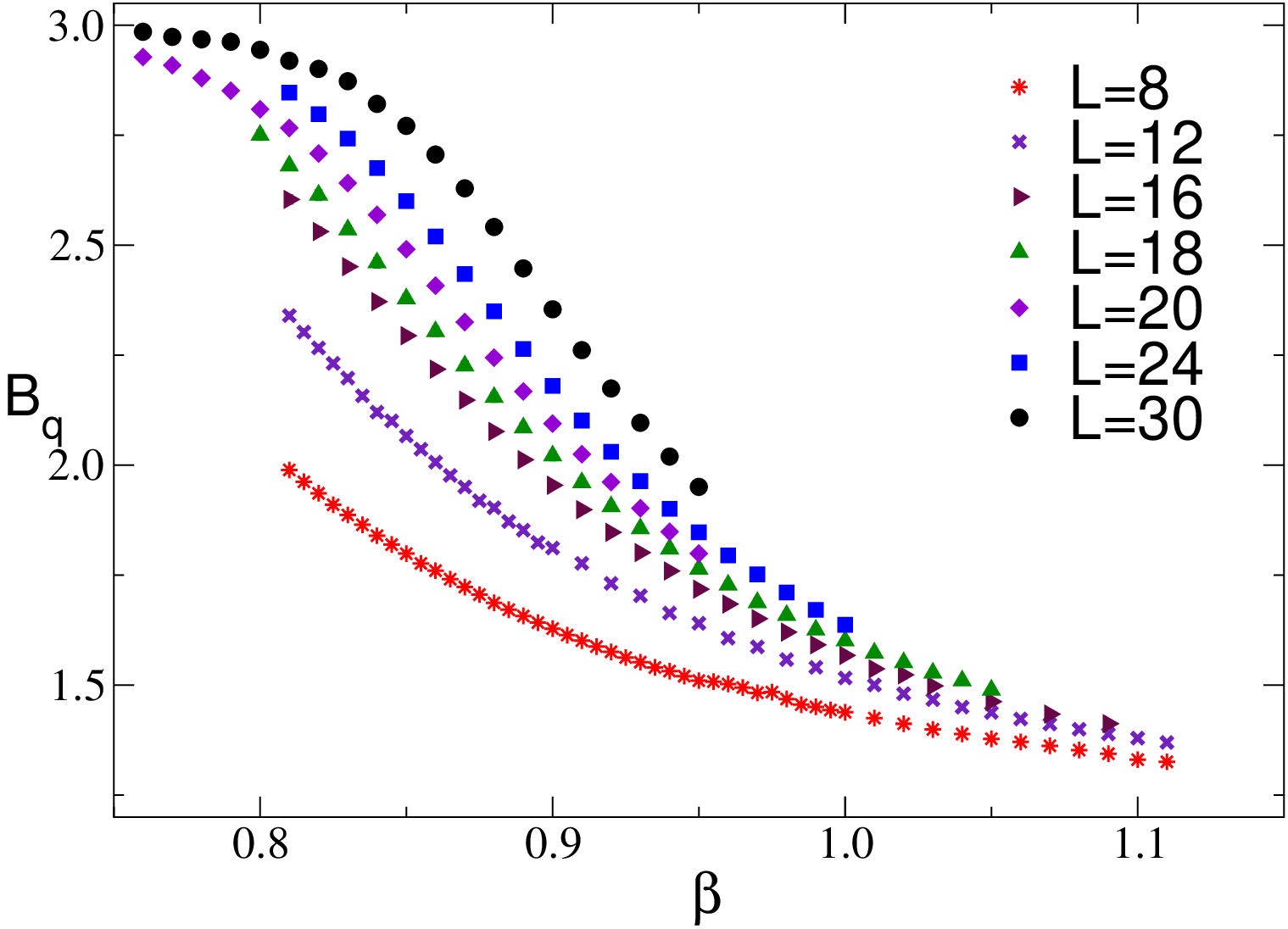}
    \setlength{\mathindent}{1em}
    \caption{The quartic cumulant $B_q$ of the overlap parameter for several lattice sizes.}
    \label{biq}
  \end{minipage}
  \begin{minipage}[t]{0.5\linewidth}
    \includegraphics[width=0.9\textwidth, keepaspectratio]{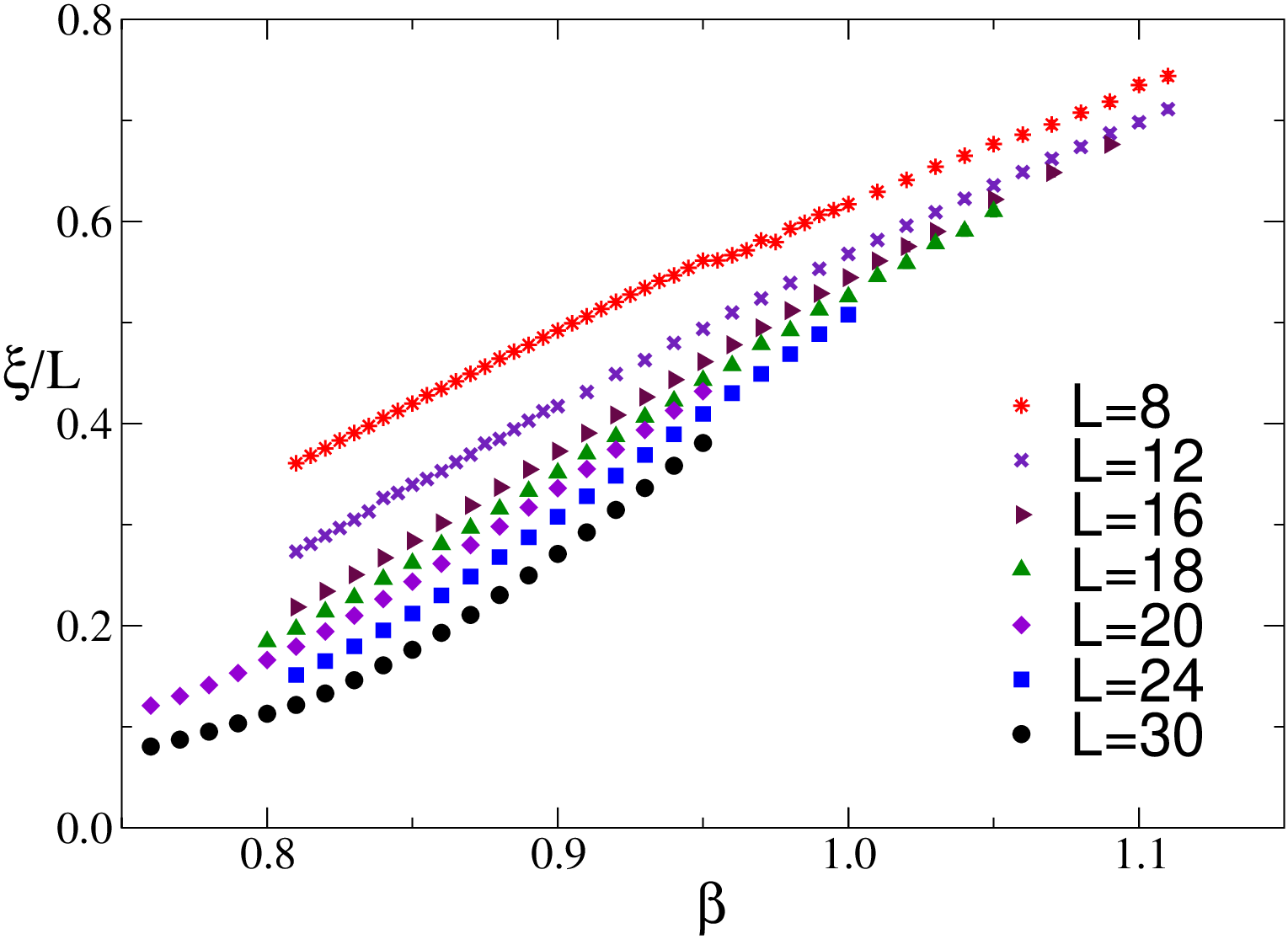}
    \setlength{\mathindent}{1em}
    \caption{The ratio $\xi/L$ for several lattice sizes.}
    \label{xil}
  \end{minipage}
\end{figure}

In MC simulations the critical point is often determined by 
considering the crossing point $\beta_{\rm cross}$ of the Binder cumulant
$B_q \equiv \overline{\mu_{4}} / \overline{\mu_{2}}^2$, 
where $\mu_{k} \equiv \langle (\sum_x q_x)^k \rangle$,
or of the ratio $\xi/L$: indeed, 
$\beta_{\rm cross}\to \beta_c$ as $L\to\infty$. 
Our data for $B_q$ and $\xi/L$, which are reported in 
Figures \ref{biq} and \ref{xil}, are compatible with
$\beta_{\rm cross} \approx 1.08$. Indeed, 
the estimates of both $B_q$ and $\xi/L$
at fixed $L$ get closer as $\beta$ increases towards 1.08,
although a crossing point is not clearly observed for those values of 
$L$ ($L\le 16$) that extend up to $\beta = 1.10$. 
This is probably due to scaling corrections that are particularly large in the
SRAM.

We can perform a
more quantitative check by using the results of \cite{KKY-06} for the
critical-point values $B_q^*$ and $(\xi/L)^*$. They quote:
$B_q^* = 1.475(6)$ (bimodal distribution) and $B_q^* = 1.480(14)$
(Gaussian distribution); $(\xi/L)^*=0.627(4)$ (bimodal distribution)
and $(\xi/L)^*=0.635(9)$ (Gaussian distribution). These results are
compatible with ours for $B_q$ and $\xi/L$ close to
$\beta = 1.08$. For $\beta = 1.07$ we have
$B_q = 1.411(4)$  ($L=12$), $B_q = 1.434(6)$ ($L=16$),  and
$\xi/L = 0.662(4)$ ($L=12$), $\xi/L = 0.648(4)$ ($L=16$).
These results are very close to the ISGM estimates and show the
correct trend. They are therefore consistent with the existence 
of a critical point at $\beta = 1.08\pm 0.04$ and with the 
conjecture that the ISGM and the SRAM belong to the same universality class.

\end{document}